\documentclass{jltp}

\usepackage{graphicx} 

\title{Energy and Structure of Hard-Sphere Bose Gases in three and
  two dimensions}

\author{F. Mazzanti$^a$, A. Polls$^b$ and A. Fabrocini$^c$}

\address{$^a$Dep. d'Electr\`onica, Univ. Ramon Llull,
  Bonanova 8, 08022 Barcelona, Spain \\
  $^b$Dep. ECM, Univ. de Barcelona,
   Diagonal 645, E-08028 Barcelona, Spain \\
  $^c$Dip. di Fisica, Univ. di Pisa and INFN, 
  Via Buonarroti,2 I--56100 Pisa, Italy}

\runninghead{F. Mazzanti, A. Polls, and A. Fabrocini}{Energy and
  Structure of 3D and 2D Bose Gases}

\begin{document}

\maketitle

\begin{abstract}
The energy and structure of dilute gases of hard spheres in three
dimensions is discussed, together with some aspects of the
corresponding 2D systems. A variational approach in the framework of
the Hypernetted Chain Equations (HNC) is used starting from a Jastrow
wavefunction that is optimized to produce the best two--body
correlation factor with the appropriate long range. Relevant
quantities describing static properties of the system are studied as a
function of the gas parameter $x=\rho a^d$ where $\rho$, $a$ and $d$
are the density, $s$--wave scattering length of the 
potential and dimensionality of the space, respectively.  The
occurrence of a maximum in the radial distribution function and in the
momentum distribution is a natural effect of the correlations when $x$
increases. Some aspects of the asymptotic behavior of the functions
characterizing the structure of the systems are also investigated.

\end{abstract}

\section{INTRODUCTION}

The study of dilute systems has become a subject of major interest
since the achievement of Bose--Einstein condensates in low--density
atomic gases confined in harmonic traps. The gas parameter $x=\rho
a^d$ where $\rho$, $a$ and $d$ are the density, s--wave scattering
length and space dimensionality (2 or 3 in the cases under study),
govern the behavior of the gases in the dilute
regime. In the 3D case, an universal dependence of the energy on the
scattering length alone has been proved to hold up to values $x\approx
10^{-3}$, while deviations appear at higher $x$.

In three dimensions, low density expansions can be derived in the
framework of standard perturbation theories. Infinite sums of ladder
diagrams can be carried out and result in the well--known Lee and
Yang  expansion of the energy per particle~\cite{yang}
\begin{equation}
E(x) = 4\pi x \left[ 1 + {128\over 15}\sqrt{x \over \pi}
  \right] \ ,
\label{yanglee}
\end{equation}
measured in units of $\hbar^2/2ma^2$.

In two dimensions, logarithmic divergences in the scattering length
introduce additional difficulties in the derivation of similar
expressions. Still, Schick and Lieb \cite{schick,lieb} proved that at
low $x$, the energy per particle of a 2D gas of hard--core bosons
satisfies the inequality
\begin{equation}
E(x) \geq {4\pi x \over \mid\!\ln x\!\mid}\left[ 1 - {\mathcal O}(\ln
  x)^{-1/5} \right] 
\label{lieb2d}
\end{equation}
in the same units

In this work an analysis of the energy and other properties of 2D and
3D gases of Hard-Sphere bosons is presented~\cite{mazza}.  The
starting point is a variational wavefunction of the Jastrow type
\begin{equation}
\Psi_0({\bf r}_1, {\bf r}_2, \ldots, {\bf r}_n) = \prod_{1\leq i<j\leq
  N} f(r_{ij})
\label{jastrow}
\end{equation}
which is known to accurately account for most of the energy and
structure of homogeneous Bose systems at low densities. The
minimization of the energy 
corresponding to the Hard--Spheres potential
\begin{equation}
V(r)= \left\{
\begin{array}{ll}
\infty & r< a \\
0 & r> a
\end{array}
\right.
\label{Vhard}
\end{equation}
produces a suitable wavefunction that can be used to calculate the
energy as well as other, relevant quantities describing the
ground state of the system.

\section{HNC FORMALISM}

The energy per particle of an homogeneous gas of Hard--Sphere bosons
described through a Jastrow wavefunction can be expressed in terms of
the two--body correlation factor $f(r)$
\begin{equation}
E = -{1\over 2} \rho\,\int d{\bf r}\, g(r) 
{\hbar^2 \over 2m} \nabla^2 f(r) 
\label{Evar}
\end{equation}
and  the radial distribution function $g(r)$,
\[
g(r_{12}) = {N(N-1)\over\rho^2} 
{\int d{\bf r}_3 d{\bf r}_4 \cdots d{\bf r}_N  \mid\! \Psi_0 \!\mid^2 
\over 
\int d{\bf r}_1 d{\bf r}_2 \cdots d{\bf r}_N  \mid\! \Psi_0 \!\mid^2} \ .
\]

One can formally express $f(r)$ as a functional of $g(r)$ and solve
the Euler-Lagrange problem $\delta E[g]/\delta g(r)=0$. This leads to
a set of 
equations which are easily expressed in terms of the static
structure function, $S(k)=1+\rho\int d{\bf  r}\,e^{i{\bf k}\cdot{\bf r}}
(g(r)-1)$. These can be solved and yield
\begin{equation}
S(k) = {t(k) \over \sqrt{ t^2(k) + 2 t(k) V_{ph}(k)}} \ ,
\label{Sk}
\end{equation}
with $t(k)=\hbar^2 k^2/ 2m$ and $V_{ph}(k)$ the Particle-Hole
interaction~\cite{campbell}. 
In the HNC/0 scheme used throughout this work, elementary diagrams are
discarded due to their small contribution at low densities. Then, for
a system of Hard Spheres and in $r$
space
\begin{equation}
V_{ph}(r) = {\hbar^2 \over m}
\mid\!\vec\nabla \sqrt{g(r)}\!\mid^2 +
\left[ g(r)-1\right] \omega_I(r) \ ,
\label{Vph}
\end{equation}
where the $k$-space induced interaction $\omega_I(k)$ reads
\begin{equation}
\omega_I(k) = -{1\over 2}\,t(k)\,
{[2 S(k)+1][S(k)-1] \over S^2(k) } \ .
\label{inducedinter}
\end{equation}

Once with  $S(k)$ or $g(r)$ one can get $f(r)$ and compute  the
momentum distribution $n(k)$
\begin{equation}
n(k) = (2 \pi)^d \rho n_0 \,\delta\!(\vec k) + 
\int d\vec r \,e^{i\vec k\cdot\vec r} [ \rho_1(r) - \rho n_0] \ ,
\label{nk}
\end{equation}
where $n_0$ is the condensate fraction and $\rho_1(r)$ is the
one--body density matrix
\begin{equation}
\rho_1(r_{11'}) = N\,{
\int d{\bf r}_2\cdots d{\bf  r}_N \Psi_0({\bf r}_1,{\bf r}_2,\ldots,
     {\bf r}_N)
\Psi_0({\bf r}_{1'}, {\bf r}_2,\ldots, {\bf r}_N)
\over 
\int d{\bf r}_1 d{\bf r}_2 \cdots d{\bf r}_N \mid\!\Psi_0 \!\mid^2 } \ .
\label{rho1}
\end{equation}

The momentum distribution  is normalized as 
$1=\int d{\bf k}\,n(k)/(2\pi)^d\rho$ and 
its second $k$--weighted moment 
provides the kinetic
energy per particle.

Notice that for Hard Spheres the potential energy is zero
because the product $g(r) V(r)$ vanishes, and therefore the kinetic
energy coincides with the total energy of the system.

\section{RESULTS}

\begin{figure}
\centerline{\includegraphics[height=2.25in]{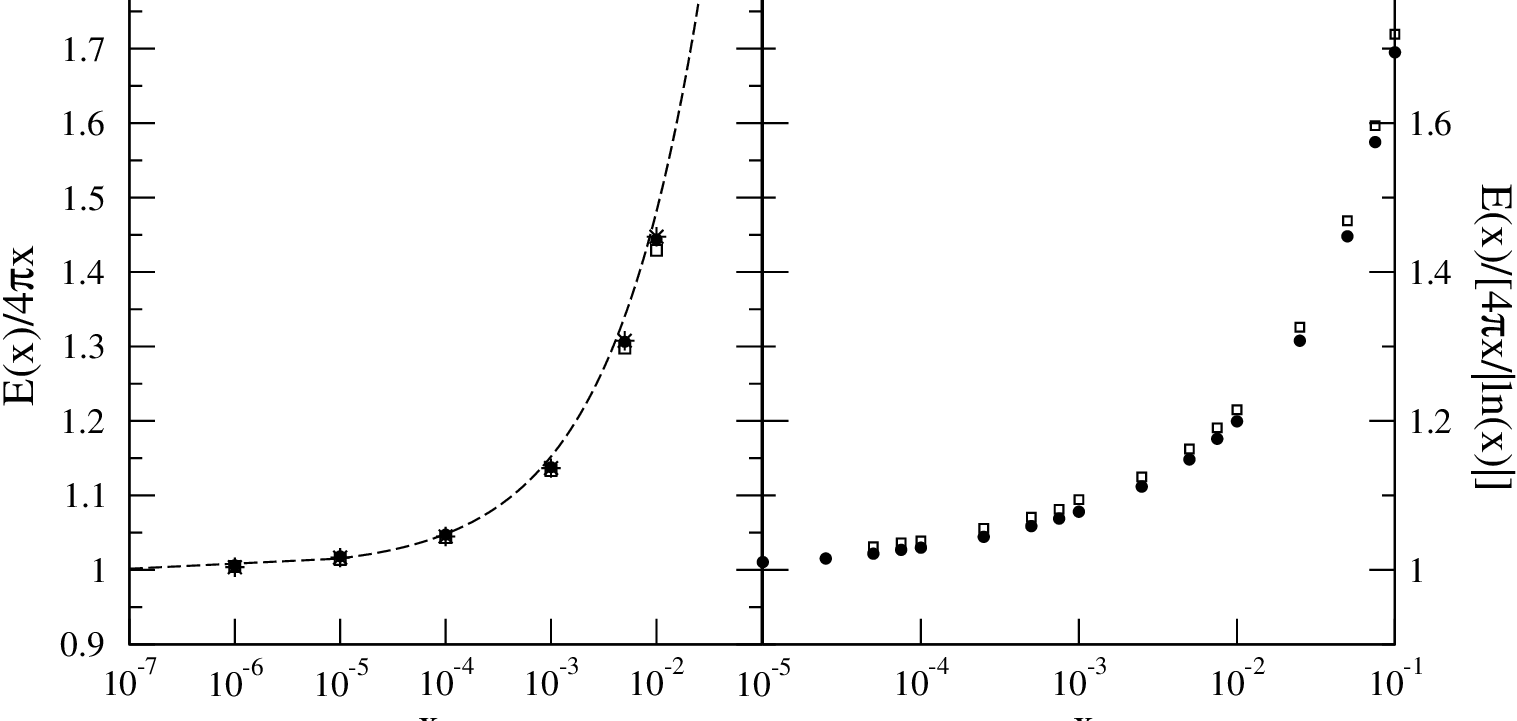}}
\caption{Energy per particle of the gas of Hard Spheres in 3D and 2D
  (left and right panels).  Solid circles, stars and open squares in
  the left stand for Euler-Lagrange, HNC/0 and Diffusion Monte Carlo
  results (the latter taken from Ref. 7. The dashed line
  represents the low--density expansion of Eq.~(\ref{yanglee}). Solid
  circles and open squares in the right are Euler--Lagrange and HNC/0
  results from the short--ranged$f_{SR}(r)$ of Eq.~(\ref{fsr2d}).}
\label{fig_E3D2D}
\end{figure}

The energy per particle of the gas of Hard Spheres gas is plotted as a
function of the gas parameter $x$ in the left panel of
Fig.~\ref{fig_E3D2D}, where the Euler--Lagrange solution is compared
with Diffusion Monte Carlo results, the low density expansion of
Eq.~(\ref{yanglee}) and the HNC prediction obtained from a
short--ranged two--body correlation function $f_{SR}(r)$. This last
function 
minimizes the lowest order in the cluster expansion of the energy of
the homogeneous gas of Hard Spheres with a healing condition at a
distance $d$, taken as a variational parameter \cite{pandha}
\begin{equation}
f_{SR}(r) = \left\{
\begin{array}{ll}
0 & r<a \\
{d\over r}\,{\sin[K(r-a)] \over \sin[K(d-a)]} & r>a
\end{array} \right. \ ,
\label{fsr3d}
\end{equation}
where $K$ fulfills the equation ${\rm cot}[K(d-a)]=(Kd)^{-1}$. As can
be seen, differences between the different approximations are not
significant up to $x\approx 0.001$. The leading term in the expansion,
$E(x)= 4\pi x$, is not enough to correctly account for the energy of
the system even at lower values of $x$, while the inclusion of the second
term, which introduces additional dependence on $x$, improves the
agreement.  The influence of the optimization on the energy is rather
small and the energy is dominated by the short range structure of the
potential.

The scaled energy per particle of the gas of Hard Spheres in 2D is
shown in right panel of Fig.~\ref{fig_E3D2D}. Euler--Lagrange results
are plotted in solid circles and compared with the energy produced by
a short range two--body correlation with a healing distance, as in the
3D case. Geometry constrains change the shape of $f_{SR}(r)$ which now
reads
\begin{equation}
f(r) = \left\{
\begin{array}{cc}
0 & r<a \\
{Y_0(\lambda a) J_0(\lambda r) - J_0(\lambda a) Y_0 (\lambda r)
\over Y_0(\lambda a) J_0(\lambda d) - J_0(\lambda a) Y_0 (\lambda d) }
& r> a 
\end{array}
\right.
\label{fsr2d}
\end{equation}
where $\lambda$ is a Lagrange multiplier satisfying the condition
$J_1(\lambda d) Y_0(\lambda a) = J_0(\lambda d) Y_1(\lambda
a)$. Energies have been divided by the leading term in
Eq.~(\ref{lieb2d}) to check convergence towards the limiting case.
The agreement of the energy with the corresponding low--density
expansion is similar to the 3D case when only the first term
$E(x)=4\pi x$ is considered, while deviations start to be significant
at $x\approx 0.001$. In any case, the singular behavior of the energy
seems to be entirely taken into account by the logarithmic term, and
thus the remaining deviations are expected to be successfully modeled
by a smooth function of $x$ that approaches zero when $x$ decreases.


\begin{figure}
\centerline{\includegraphics[height=2.25in]{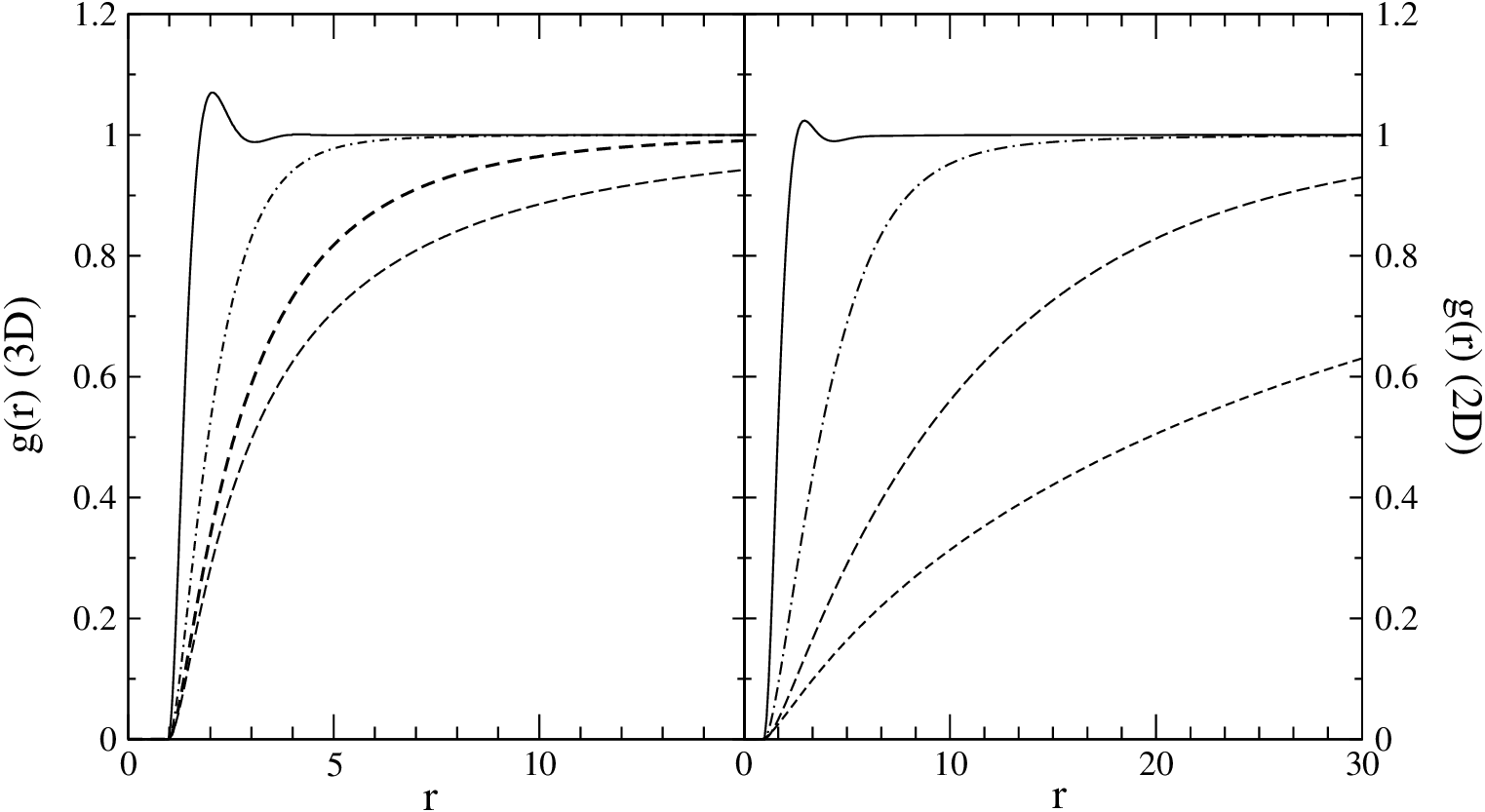}}
\caption{$g(r)$ for the 3D (left) and 2D (right) gases of Hard Spheres
  as a function of the gas parameter. Solid line: $x=10^{-1}$,
  dot--dashed line: $x=10^{-2}$, long-dashed line: $x=10^{-3}$,
  short--dashed line: $x=10^{-4}$.}
\label{fig-gr3d2d}
\end{figure}
The radial distribution functions of the 3D and 2D gases of Hard
Spheres are shown in Fig.~\ref{fig-gr3d2d}\, for several values of $x$
(left and right panels, respectively). In both cases $g(r)$ develops
a peak at $x=0.1$ that can not be resolved at lower values of $x$, a
clear signature of the effect of correlations. At low $x$'s, $g(r)$ is
a monotonically increasing function of the distance, approaching
faster and faster the asymptotic limit $g(r)\to 1$ with the density.
This is more notorious in the 2D case.  In fact, this is not
surprising if one notices that in 3D and at large distances $g(r)
\approx 1 - a/r^4$, while in 2D this behavior changes to $g(r)\approx
1 - b/r^3$, with $a,b$ constants depending on the velocity of sound in
the medium. 

\begin{figure}
\centerline{\includegraphics[height=2.25in]{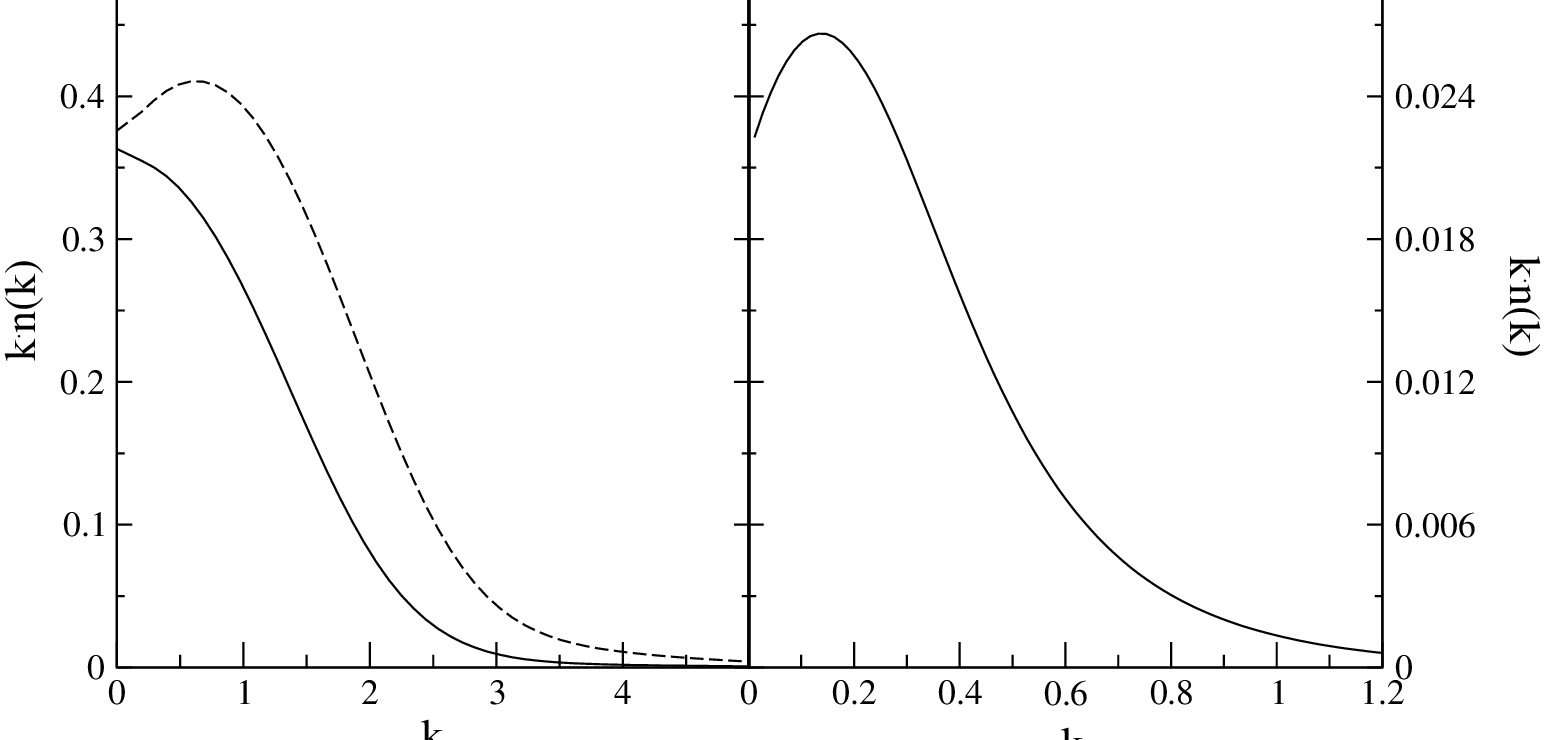}}
\caption{Momentum distribution of Hard Spheres in 3D and 2D (left and
  right, respectively). Solid and dashed lines in the left, $x=0.05$
  and $x=0.08$. Solid line in the right, $x=0.01$.}
\label{fig-nk3d2d}
\end{figure}

The final quantity analyzed is the momentum distribution. The product
$kn(k)$ is shown for the 3D and the 2D cases in
Fig.~\ref{fig-nk3d2d}. At low $x$, the momentum distribution is a
monotonically decreasing function of $k$, a behavior already shown by
the Bogoliubov approximation at any density. When $x$ increases, $k
n(k)$ develops a peak at low $k$, which can be taken as a genuine
effect of long--range correlations. This peak becomes more pronounced
and shifts to the right with increasing $x$. While this maximum is
clearly visible already at $x=0.01$ in 2D, it only shows up at
$x>0.05$ in 3D. As far as the condensate fraction $n_0$ is concerned,
the Bogoliubov estimation in 3D, $n_0^B = 1 - (8/3)\sqrt{x/\pi}$,
predicts values that are slightly higher than the ones obtained from
the solution of the Euler--Lagrange problem, yielding $n_0^B=0.850$
against $n_0=0.801$ at $x=0.01$.  Larger discrepancies arise at higher
values of $x$. Similar results hold when the EL condensate fraction is
compared with Schick's prediction, $n_0^S=1+1/\ln(x)$.


\end{document}